# Microscope Projection Photolithography Based on Liquid Crystal Microdisplay

Young-Gu Ju[1], Hyeon-Jun Kim[2] and Young-Min Ko[2]

[1]Department of Physics Education, Kyungpook National University, 80 Daehak-ro, Buk-gu, Daegu 41566, Korea

[2]Daegu Science High School, 154, Dongdaegu-ro, Suseong-gu, Daegu 42110, Korea

E-mail: ygju@knu.ac.kr

Abstract

We developed a microdisplay-based microscope projection photolithography (MDMPP) technique in which a liquid crystal (LC) microdisplay is used as a reconfigurable photomask for a microscope projector. The LC microdisplay provides a significant advantage in terms of cost and speed since patterns can be generated through software instead of redesigning and fabricating glass photomasks. The constructed MDMPP system could produce line patterns as narrow as 2.4 μm, smaller than that specified by the diffraction limit, with the aid of a 4X objective lens. The achievement of a linewidth smaller than the theoretical limit may be ascribed to a combination of overexposure and the underetching effect, in addition to the good optical performance of the system. In a diffraction experiment performed with fabricated slits, the application of the MDMPP technique helped provide various patterns of the slits, demonstrating the potential usefulness of the MDMPP system in undergraduate optics courses. We expect that MDMPP can contribute to the field of physics education and various areas of research, such as chemistry and biology, in the future.

## 1. Introduction

Photolithography is one of the major techniques used in the fabrication of integrated circuits [1]. Currently, it is being used in many research fields, such as microelectromechanical systems, microfluidics, biosensors, biology, chemistry, physics, and material science, for generating small patterned microstructures. Consequently, many academic institutes possess mask aligners that enable the transfer of a photomask pattern to a substrate coated with photoresist (PR). The mask aligners used in academic institutes are mostly of the contact printing type, which are used for relatively low-resolution applications. Despite not having a very high resolution, they are costly and bulky since they employ a high-power mercury lamp with a large cooling system. Furthermore, a contact mask aligner requires a fine-patterned glass photomask. Since the pattern on the photomask is directly transferred onto the substrate, the mask must have the same resolution or precision as the final pattern on the substrate. Since the use of a high-resolution photomask requires the use of e-beam lithography or laser direct writing on a chrome mask, it is very time-consuming and expensive to fabricate.

In a previous study [2], we suggested that microscope projection photolithography (MPP) [3–6] based on an ultraviolet (UV) light-emitting diode (LED) could replace a mask aligner equipped with a mercury lamp, providing several advantages in terms of the cost and size of equipment. UV-LED-based MPP can produce line patterns as wide as 5 μm under optimal lithography conditions. Although the study demonstrated the usefulness of a microscope projector in the field of education, there were problems in fabricating a precision photomask with the same precision as the target pattern. For an example, if the final pattern had a linewidth of 2 μm, then the photomask should have the same linewidth. In general, such a precision photomask can be obtained only by using e-beam lithography and a laser writer, not with conventional equipment such as a laser printer or an inkjet printer, resulting in high cost and a long experiment preparation time. Another problem with this glass photomask is that changing the mask pattern involves redesigning the pattern using computer-aided design (CAD) and fabricating a new photomask.

In this paper, we propose MPP based on a liquid crystal (LC) microdisplay. Apart from the use of a UV LED as a light source, we replaced the glass photomask with an LC microdisplay for the easy reconfiguration of photomask patterns by using software. An LC microdisplay is a small display device that is usually used in an LC projector. A pixel of the LC microdisplay is as large as several microns, and therefore, it can be reduced to an image of few microns by the objective lens in MPP. If the glass photomask used in MPP is replaced by an LC microdisplay, the photomask pattern can be easily modified by sending graphic data on the new photomask pattern to the display driver. We examined the feasibility of the LC microdisplay as a reconfigurable photomask in an MPP system, and its use solved the aforementioned problems associated with a glass photomask. The LC microdisplay considerably reduced the time and cost involved in changing the mask pattern.

Here, we describe how we constructed the system for microdisplay-based MPP (MDMPP) and how we checked its performance limit by preparing various aperture patterns and optimising the photolithography conditions. Furthermore, we present the results of a diffraction experiment involving the use of slits of various shapes and sizes fabricated using the MDMPP system. We expect that this study can help undergraduate students build their own lithography system and fabricate apertures of different sizes and shapes in the laboratory, which would help increase their understanding of the diffraction phenomenon. This MDMPP system can also serve as a useful laboratory instrument in many other fields requiring microstructure patterning, such as physics, chemistry, mechanics, and electronics.

## 2. Description of the method

In this section, we describe the construction of the proposed MDMPP system, the establishment of photolithography conditions, and diffraction experiments conducted with fabricated slits. The photolithography process includes the coating of PR on a glass substrate, fabrication of photomasks, UV exposure of the substrate, development of PR patterns, chemical etching, and removal of the PR. The photolithography process and its optimisation are similar to those of a previous study [2].

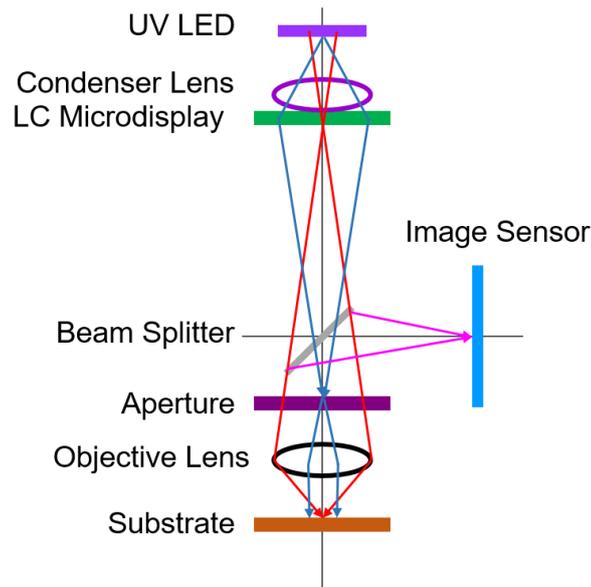

Figure 1. Schematic of the MPP system based on an LC microdisplay.

MDMPP is very similar to MPP, proposed in a previous study [2], except for the photomask being replaced with an LC microdisplay. The schematic in figure 1 shows the optical setup and the related major rays. The red rays cross the optical axis at the substrate, photomask, and image sensor, and these crossing points are conjugate points. This implies that an image of the pattern generated on the LC microdisplay is formed on both the substrate and image sensor. The image sensor can be used to confirm the correct alignment of the substrate surface with the UV image of the LC microdisplay. On the other hand, the blue ray crosses the optical axis at the UV LED and aperture stop. Therefore, the image of the LED is formed on the aperture stop, and the blue ray illuminates the substrate uniformly since the plane after the condenser lens is conjugate to the substrate according to the Köhler illumination concept [7]. While we modified a conventional microscope into an MPP system in a previous study [2], we constructed a microscope structure in this study by using an optical cage system, as shown in figure 2. The construction of a new microscope provided better controllability for all the optics involved compared to the modification, and consequently, the system could be easily redesigned and modified if required. The four cylindrical rods formed rails, and optical mounts were introduced to position and fix optical components. Most of the mounts were made using a 3D printer, to reduce the cost. The sample substrate was attached to the XYZ stage to enable the fine adjustment of the sample to match the positions of the sample and the LC microdisplay image. The MDMPP system contained a 4X objective lens with a numerical aperture (NA) of 0.10. Owing to the lens magnification, the pattern on the microdisplay was reduced by about four times when it was reproduced on the substrate. Since the actual tube length was close to (but not exactly) 160 mm, the magnification factor of four was an approximate value.

Figure 2(b) shows a photograph of the MDMPP system taken during its use, and it shows the light from the LED. An image captured by the image sensor is shown in figure 2(c), and it confirms the correct alignment of the setup. We tested the system by placing a silver-coated glass plate at the substrate position. The scratch pattern on the silver coating was well aligned with the arrow pattern originating from the LC microdisplay, indicating that the LC microdisplay, substrate, and image sensor were conjugate to each other. The power consumption of the UV LED used in the MDMPP system was 1 W and its wavelength was around 390 nm.

The LC microdisplay used in the system was EPSON L3C07U-85, and it was purchased from bbs bild- und lichtsysteme [8]. The panel had 1920 × 1080 pixels and dimensions of 16.32 mm × 9.18 mm. The size of each pixel was 8.5 ㎛ × 8.5 ㎛. The driving board was necessary to display the image of the notebook on the microdisplay and it was purchased together with the microdisplay.

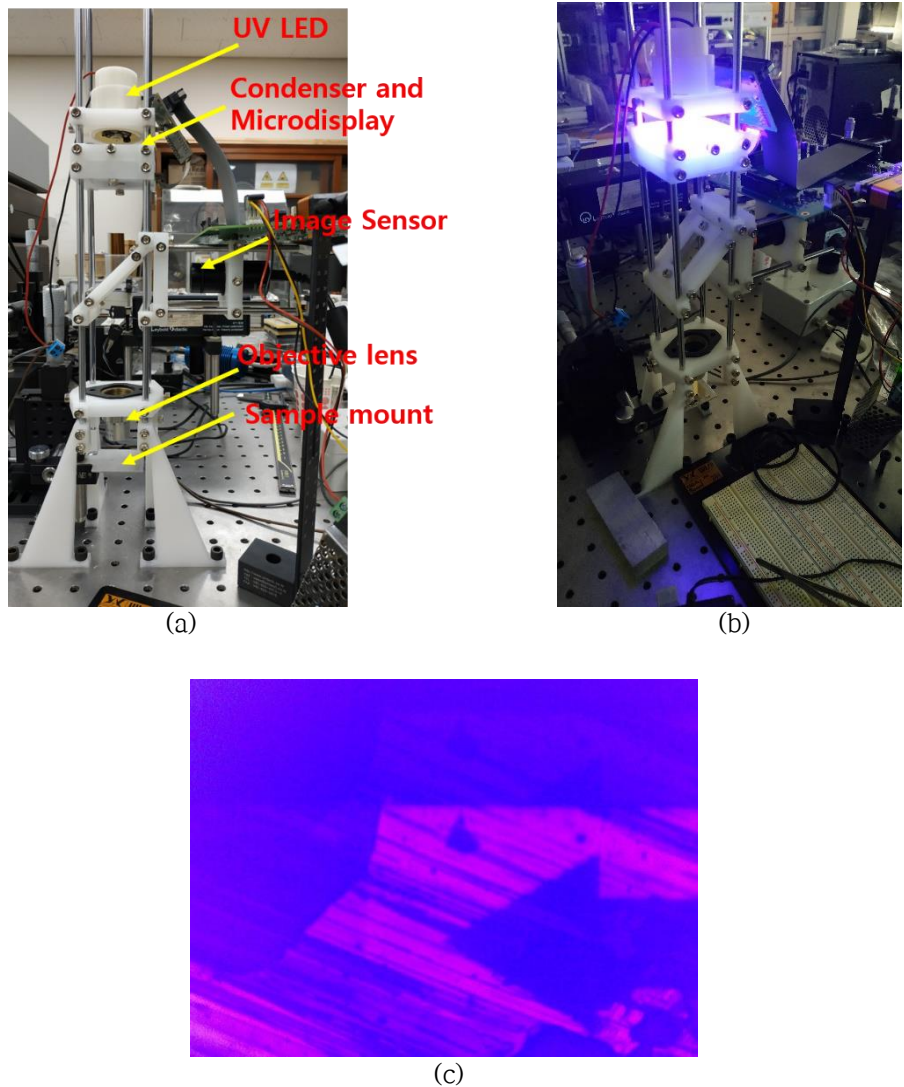

Figure 2. Assembled MDMPP system and the image of a substrate captured by it. (a) The assembled optical system for MDMPP, (b) the system with the UV LED turned on, and (c) the image of a substrate captured by the image sensor on the side. The image in (c) confirmed the alignment between the arrow and the scratches, which corresponded to the pattern on the microdisplay and that on the silver-coated surface of the test substrate, respectively.

The MDMPP process started with the coating of silver onto a glass plate. A sputtering machine evaporated silver onto a glass plate; the thickness of the silver layer was estimated to be about 200 nm. The silver-coated glass plate was spin-coated with PR at 3000 rpm for 30 s. The PR was then soft-baked on a hot plate at 100 ℃ for 1 min. The PR used in this study was obtained by thinning AZ4562 PR with 2-methoxy-1-methylethylacetate.

The second step of this study was to prepare a photomask pattern. The fabrication of a photomask was not required, unlike conventional photolithography, since the required photomask pattern could be generated on the LC microdisplay through programming. Figure 3 shows an example of a code and the double-slit pattern generated by it. When this pattern was displayed on the notebook screen, the image was transferred to the driving board and the microdisplay. Although any software such as PowerPoint can generate patterns on a notebook screen and hence on a microdisplay, we used Python codes to control the pattern with an accuracy of a pixel. Thus, we could generate various patterns easily without fabricating photomasks. In the test run performed for process optimisation, we used PowerPoint too to generate simple patterns such as rectangles.

In order to obtain high-contrast patterns in MPP, we focused on the optimisation of the UV exposure time and developing time. After UV exposure, the sample was developed with the AZ400 developer for 1–3 min. The developing time and UV exposure time were varied to achieve the best resolution for the PR pattern. UV exposure typically took 10–20 min, depending on the pattern. Thick PR layers and patterns smaller than 5 μm required a longer exposure time for good patterning. The exposure time was considerably longer than that in the previous MPP system [2], and it could be attributed to the polariser absorbing a large percentage of the UV radiation.

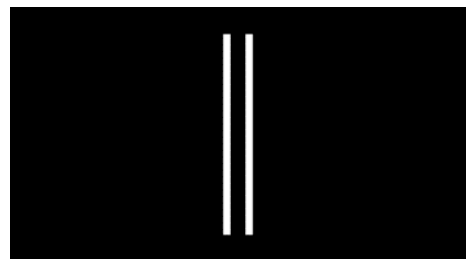

(a)          (b)

Figure 3. (a) Example Python code for generating a photomask pattern and (b) the pattern generated on the microdisplay by the code.

The etching of the silver coating followed the development process. The etching solution for the silver was prepared by mixing 100 mL of water, 4.15 mL of ammonium hydroxide, and 4.13 mL of hydrogen peroxide [9]. The sample with the PR pattern was immersed in the etching solution, which was then stirred slowly to promote the diffusion of the etching chemicals onto the sample surface. After the etching, the sample was rinsed with water, dried using a nitrogen gun, and inspected under a microscope for evaluating the pattern quality. The lithography conditions were optimised by repeating the procedure with different exposure times, developing times, and etching times until a good quality pattern was obtained.

After a good quality pattern was obtained, the PR on the substrate was removed by spraying acetone.

The use of the MDMPP system considerably reduced the process optimisation time, since we could control the mask patterns with software. For example, when we tested the UV exposure conditions, we drew six rectangles on a PowerPoint page and copied them to five more pages. As the rectangles were successively copied to each of the five pages, we removed a specific number of rectangles from the page, the number being the page number (1–5) minus one. If we displayed each slide for a certain duration, each rectangle had a different exposure time. Consequently, one sample had six rectangular patterns with six different UV exposure times. In this manner, we could test six different exposure times simultaneously.

The microscope's digital camera recorded an image of each sample for the evaluation of the lithography conditions. The microscope images were calibrated using a microscope image of a ruler, shown in figure 4, with ten 10 micron markings per division. All the microscope images were recorded for the same magnification of the objective lens and with the same video frame size (800 pixels × 600 pixels). The conversion rate from pixels to microns was about 0.673 $\mu$m/px, and it was approximately the same in both horizontal and vertical directions. The quality of lithography was chiefly assessed by checking the linewidths and the quality of the slit pattern after development, etching, and PR removal.

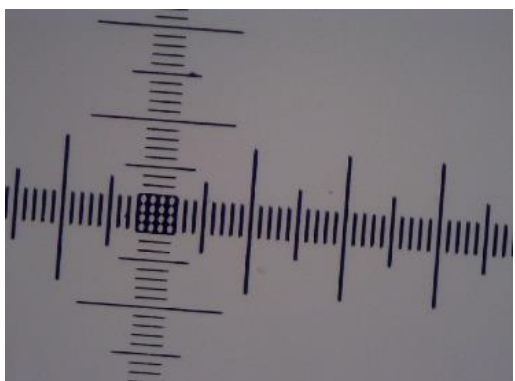

Figure 4. Microscope ruler used for image calibration. The smallest division is 10 $\mu$m.

## 3. Results and discussion

We fabricated various types of aperture patterns, as shown in figure 5, using the MDMPP system and the lithography process. The pattern sizes were expressed in 'mpx' as well as in $\mu$m. The unit 'mpx' represents a pixel in the microdisplay. The length in mpx was not proportional to the length in $\mu$m because the etching process usually changes the size of the PR pattern. Usually, overetching or undercut etching widens the etched area, while underetching does the opposite. Therefore, the actual size of the etched pattern depended on both the lithography process and etching conditions. Nevertheless, the ratios of the pattern size in micron to the pattern size in mpx for figure 5(b), (c), and (d) are roughly 2 $\mu$m/mpx, which could also be obtained by dividing the pixel size (8.5 $\mu$m) by the magnification factor ($\approx$X4) of the objective lens. In the case of figure 5(a), overetching could widen the etched area and reduce the inner radius of the ring, leading to a smaller size than that expected from the conversion ratio. The fabrication of the various patterns demonstrated that the MDMPP system could fabricate microscale patterns through computer-generated images more easily compared to the use of the conventional glass photomask.

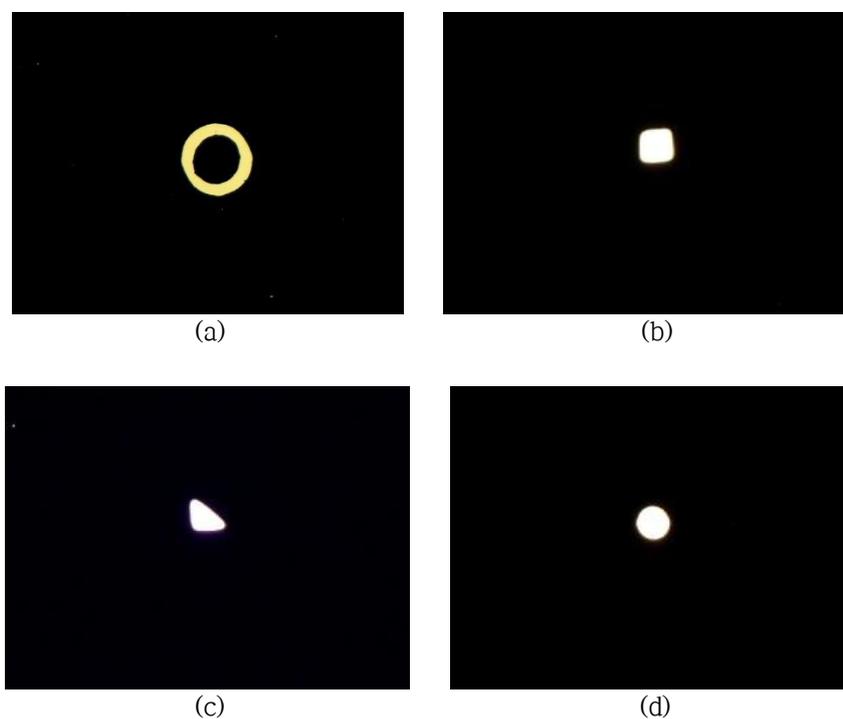

(a) (b)

(c) (d)

Figure 5. Transmission microscope images of various patterns produced using the lithography system: (a) a ring with an inner diameter of 40 mpx (≈65 μm) and an outer diameter of 50 mpx (≈ 98 μm), (b) a square with an edge of 20 mpx (≈42 μm), (c) a right triangle with a base of 20 mpx (≈46 μm), and (d) a circle with a diameter of 20 mpx (≈44 μm). The unit 'mpx' represents a pixel on the microdisplay.

The second set of patterns we tested comprised double slits with various widths and centre-to-centre separations, as shown in figure 6. The double-slit experiment is one of the basic physics experiments that often appears in optics textbooks, and it explains the principle of superposition of waves. Although a single slit can be easily formed either by employing the doctor blade method [10, 11] or by using an old method involving Aquadag and a blunt pin [12], it is not easy to control the width and separation of a double slit. Most of the kits used for the diffraction experiment include a predefined set of slits with several combinations of the width and separation. In this study, we tested the capability of MDMPP by fabricating various double slits. Each double slit is displayed along with the diffraction pattern and its intensity profile in figure 6.

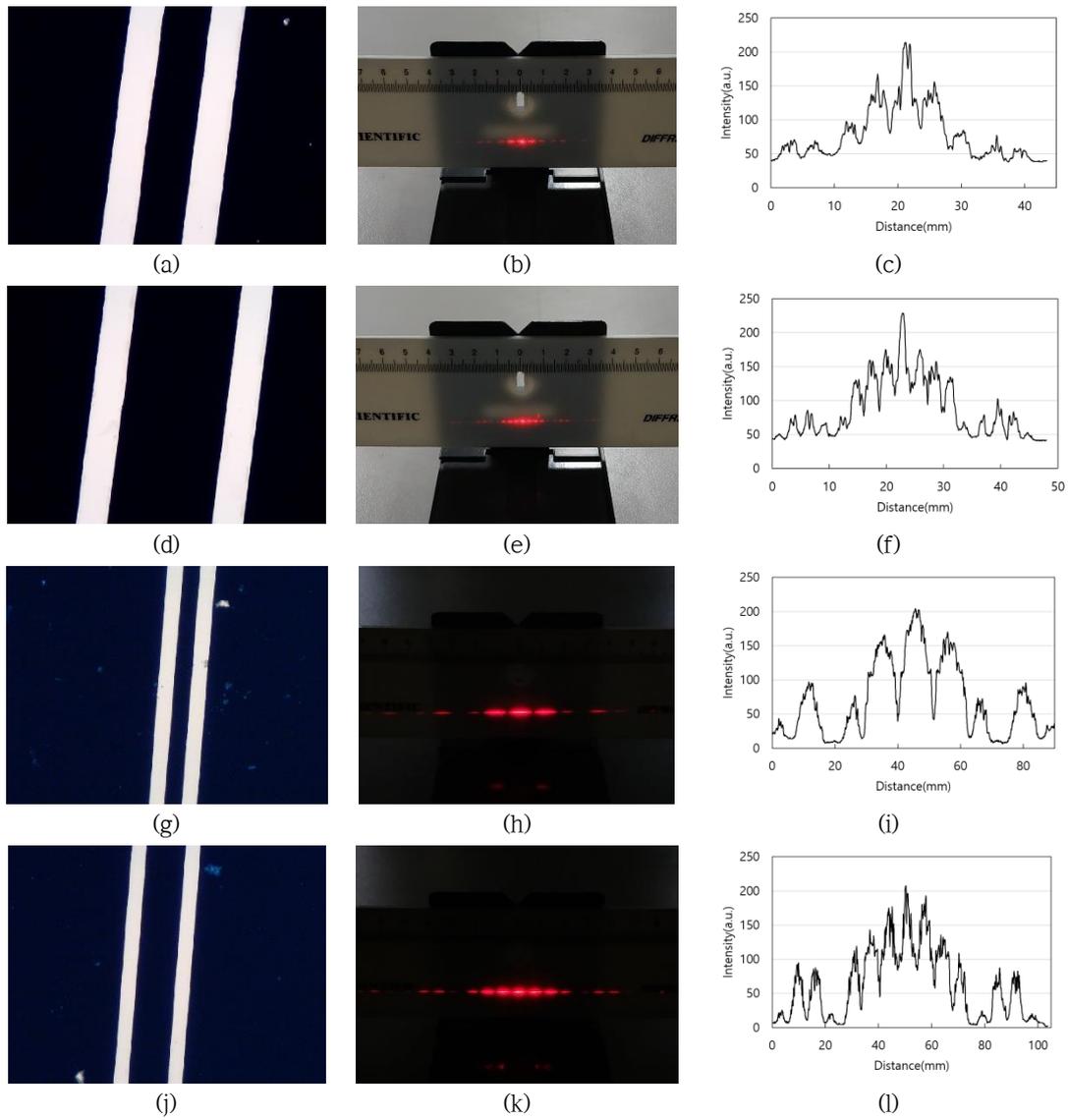

Figure 6. Double slits fabricated by the lithography system. The diffraction pattern and the intensity plot of a slit are displayed in the same row as the slit image: (a) width = 25 mpx, separation = 75 mpx; (d) width = 25 mpx, separation = 125 mpx; (g) width = 10 mpx, separation = 30 mpx; and (j) width = 10 mpx, separation = 50 mpx. 'Separation' refers to the centre-to-centre distance between the two slits.

The microscope images of the double slits were analysed to measure the width and separation, and the results are presented in table 1. The separation of a double slit obtained from microscope measurements and the distance between the intensity peaks of the diffraction patterns could be used to estimate the wavelength of the laser used in this study as follows:

$$a \sin \theta = \lambda \qquad (1)$$

$$a \frac{d}{L} \approx \lambda \qquad (2)$$

Here, $a$, $\theta$, $\lambda$, $d$, and $L$ represent the slit width, diffraction angle, wavelength, distance between adjacent intensity peaks, and the distance between the slit and the screen, respectively. The intensity profile of the diffraction pattern was extracted from figure 6(b), (e), (h), and (k) by using ImageJ software [13] to measure the distance between two maxima; the intensity profiles are shown in figure 6(c), (f), (i), and (l). We determined the conversion factor between the image pixel and the real dimension of the diffraction patterns by measuring a known length of the ruler shown in these images in units of image pixel. $L$ was 1000 mm in this study. The wavelength estimated from each diffraction pattern is shown in table 2. The laser wavelength was known to be 650 nm from the data provided by the manufacturer. The relative error of the estimated wavelength was less than 3%, which was within the measurement error of the diffraction experiment.

Table 1. Dimensions of the double slits obtained by analysing the images in figure 6. 'Length (px)' indicates measurements in units of the pixel of the image.

| Slit | Part | Length (mpx) | Length (px) | Length (μm) |
|---|---|---|---|---|
| (a) | Separation | 75 | 204 | 137 |
|  | Width | 25 | 84 | 57 |
| (d) | Separation | 125 | 337 | 227 |
|  | Width | 25 | 79 | 53 |
| (g) | Separation | 30 | 81 | 55 |
|  | Width | 10 | 39 | 26 |
| (j) | Separation | 50 | 136 | 92 |
|  | Width | 10 | 38 | 26 |

Table 2. Analysis of the diffraction patterns of the double slits from the images in figure 6.

| Slit | $d$ (mm) | $a$ (μm) | $\lambda$ (nm) from diffraction pattern |
|---|---|---|---|
| (a) | 4.8 | 137 | 660 |
| (d) | 2.8 | 227 | 640 |
| (g) | 11.4 | 55 | 630 |
| (j) | 7.0 | 92 | 640 |

In the third experiment, the MDMPP system produced multiple slits, which seemed too difficult to produce by the traditional doctor blade [10, 11] or Aquadag method [12]. The slit width was 10 μm and the separation values were 30 and 50 μm. The number of slits was 10 and 30. The fabricated aperture patterns, corresponding diffraction patterns, and cross-sectional profiles are shown in figure 7. The diffraction patterns of the multiple slits in figure 7 show sharper peaks than those of the double slits in figure 6. According to diffraction theory, this overall change in the diffraction patterns results from the increased number of slits in the aperture. Similar to the double slit patterns, the dimensions and diffraction patterns of the multiple slits were analysed to estimate the laser wavelength, and the results are presented in tables 3 and 4. The relative error of the wavelength was less than 1.6%, which was reasonable considering the measurement error. In this way, we demonstrated that MDMPP could produce various aperture patterns that could be useful for teaching diffraction experiments.

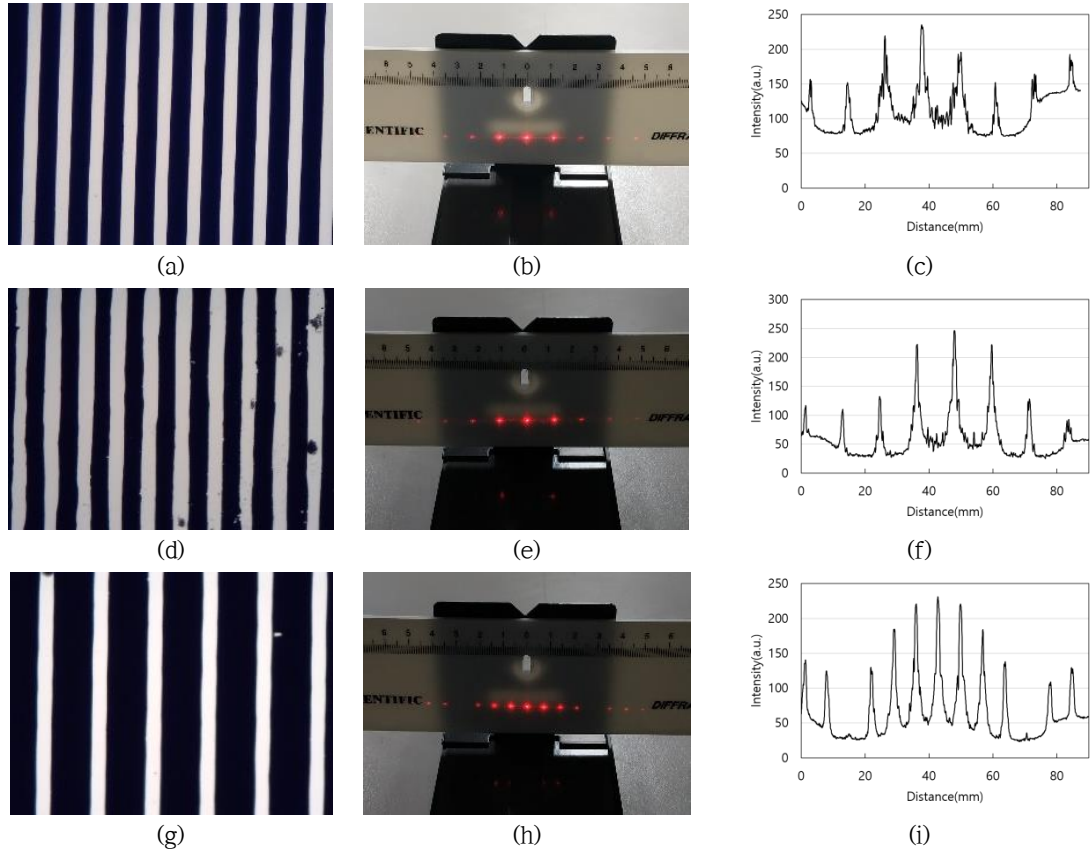

Figure 7. Multiple-slit patterns fabricated by MDMPP: (a) width =10 mpx, separation = 30 mpx, number of slits = 10; (d) width = 10 mpx, separation = 30 mpx, number of slits = 30; and (g) width =10 mpx, separation = 50 mpx, number of slits = 30. The diffraction patterns of the slits are shown by their side (b, e, and h), and the intensity profiles of the diffraction patterns (c, f, and i) are shown next to the diffraction patterns.

Table 3. Dimensions of the multiple slits obtained by analysing the images in figure 7.

| Slit | Part | Length (mpx) | Number of slits | Length (px) | Length (μm) |
|---|---|---|---|---|---|
| (a) | Separation | 30 | 10 | 83 | 56 |
|     | Width      | 10 |    | 34 | 23 |
| (d) | Separation | 30 | 30 | 82 | 55 |
|     | Width      | 10 |    | 35 | 24 |
| (g) | Separation | 50 | 30 | 136 | 92 |
|     | Width      | 10 |    | 38 | 26 |

Table 4. Analysis of the diffraction patterns of the multiple slits from the images in figure 7.

| Slit | d (mm) | a (μm) | λ (nm) from diffraction pattern |
|---|---|---|---|
| (a) | 11.6 | 56 | 650 |
| (d) | 11.8 | 55 | 650 |
| (g) | 7.0 | 92 | 640 |

Finally, we attempted to produce the smallest possible linewidth of a pattern with the MDMPP system. The experimental procedure was similar to that for the fabrication of multiple slits, except that the width was 1 or 2 mpx and there were 10 evenly spaced slits in a pattern. The results are shown in figure 8 and table 5. In order to obtain a fine line pattern, we increased the UV exposure time and reduced the etching time. Generally, a small linewidth reduces the amount of light arriving at the PR, and therefore, increased UV exposure was necessary to achieve a small linewidth. Underetching prevented undercut etching, which could otherwise cause an unnecessary widening of the line pattern. The width of the two fabricated multiple slits were 2.4 and 7.4 μm. Furthermore, 1 mpx corresponded to 1.8 μm according to the data presented in figures 6 and 7. This conversion factor was usually valid for the separation, while the width seemed to be affected by the etching conditions and other process parameters. Notably, the linewidth achieved was quite small considering the diffraction limit. Since a 4X objective lens was used in the system, the diffraction theory gives the resolution as follows [14]:

$$\text{Resolution} = 1.22\,\lambda \times f/\# = 1.22 \times 0.39\,\mu m \times 5.0 = 2.4\,\mu m \quad (3)$$

$$f/\# \approx \frac{1}{2\,N.A.} = \frac{1}{2\,(0.10)} = 5.0 \quad (4)$$

Here, f/# and N.A. denote the f-number and the NA of the objective lens. The geometric size of a pixel in the microdisplay was 1.8 μm according to the conversion factor between the image pixel and the real dimension. If the half maximum intensity of the diffraction pattern was effective for the exposure of the PR, the resolution should be added to the geometric size of a pixel. Therefore, the theoretical linewidth limit is about 4.2 μm, which is larger than that obtained in the experiment. This implies that the processing conditions such as overexposure and underetching may have reduced the linewidth of the etched pattern beyond the diffraction limit. The wavelengths estimated from the diffraction patterns are shown in table 6, and they agree well with the original values provided by the manufacturer.

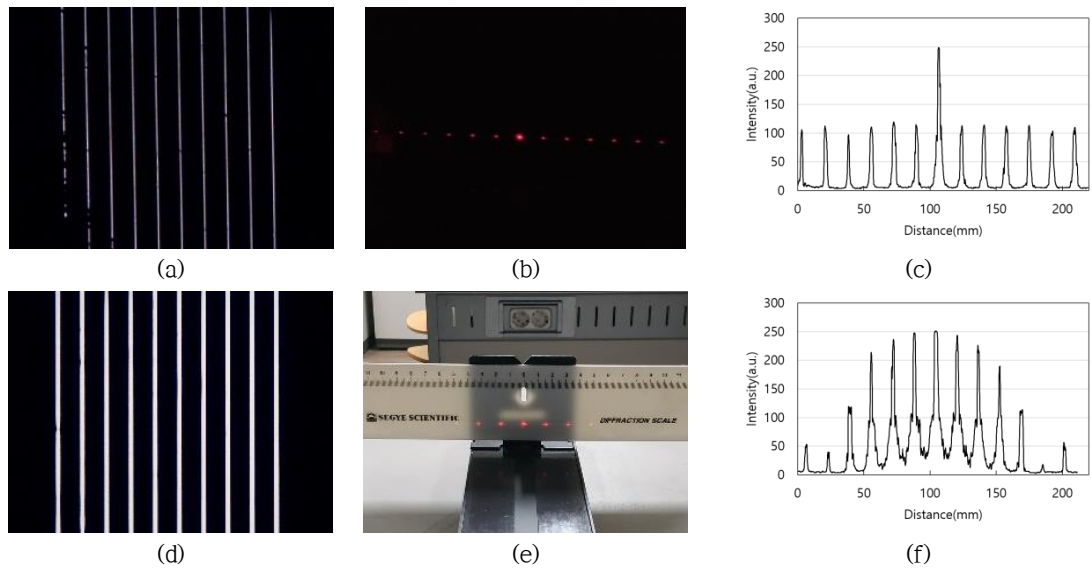

Figure 8. Demonstration of narrow linewidth formation by the MDMPP system: (a) Slits with width = 1 mpx and separation = 21 mpx, (b) the diffraction pattern of the slit in (a), and (c) the intensity profile of the diffraction pattern in (b). (d) Slits with width = 2 mpx and separation = 22 mpx, (e) the diffraction pattern of the slits in (d), and (f) the intensity profile of the diffraction pattern in (e).

Table 5. Dimensions of the multiple slits obtained by analysing the images in figure 8.

| Slit | Part | Length (mpx) | Number of slits | Length (px) | Length (μm) |
|---|---|---|---|---|---|
| (a) | Separation | 21 | 10 | 57 | 38 |
|  | Width | 1 |  | 3.6 | 2.4 |
| (d) | Separation | 22 | 10 | 60 | 40 |
|  | Width | 2 |  | 11 | 7.4 |

Table 6. Analysis of the diffraction patterns of the multiple slits from the images in figure 8. The wavelength was obtained from the diffraction patterns to estimate the error.

| Slit | $d$ (mm) | $a$ (μm) | $\lambda$ (nm) from diffraction pattern |
|---|---|---|---|
| (a) | 17.1 | 38 | 650 |
| (d) | 16.2 | 40 | 650 |

## 4. Conclusions

We constructed an MDMPP system in which an LC microdisplay was used as a reconfigurable photomask for a microscope projector. The use of the LC microdisplay improved MPP by offering a considerable advantage in terms of pattern generation: the photomask pattern could be changed through software and a glass photomask was not required. The reconfigurable capability of the LC photomask saves considerable time and cost, especially when various patterns are required for experimental purposes. The MDMPP system was built with the skeleton of a 3D-printed optical cage system, and it comprised a reflection optical microscope system with UV LED illumination and an LC microdisplay photomask.

The constructed MDMPP system produced various micropatterns such as a ring, a rectangle, a triangle, and a circle with dimensions as large as 40–60 μm. Furthermore, it successfully fabricated double slits and multiple slits with various widths and separations. We measured the dimensions of the fabricated slits and imaged the diffraction patterns. The wavelength estimated from the diffraction patterns of the slits agreed well with that provided by the manufacturer within the measurement error. The diffraction experiments conducted with fabricated slits in this study are examples of experiments in optics courses for which the MDMPP setup could be used.

Finally, we determined the linewidth limit of the MDMPP system by fabricating multiple slits with a width equal to a pixel of the microdisplay. The linewidth of the final pattern formed by the 4X objective lens was 2.4 μm, which was smaller than that specified by the diffraction limit. The achievement of a linewidth smaller than the theoretical limit may be attributed to a combination of overexposure and the underetching effect, in addition to the good optical performance of the system.

In summary, using the MDMPP system, we fabricated various patterns such as double slits and multiple slits with different widths and spacing, which are usually difficult to produce in the laboratory without employing photolithography techniques. In the diffraction experiment involving fabricated slits, the diffraction patterns of the various slits were clearly observed. This demonstrates the potential usefulness of the MDMPP system in teaching optics at the undergraduate level. We expect that MDMPP contribute to the field of physics education and other areas of research, such as chemistry and biology, in the future.

## Acknowledgements
We appreciate the help provided by Haneol Park at the beginning of this study.